\documentclass{article}
\usepackage{preamble}

\title{ Artistic Strategies to Guide Neural Networks
\vspace{-0.1cm}
}

\author{
Varvara Guljajeva*‡\textsuperscript{1}, 
Mar Canet Solà*‡\textsuperscript{2},  
Isaac Joseph Clarke\textsuperscript{1}\\
\small \textsuperscript{1}The Hong Kong University of Science and Technology. Guangzhou, China\\
\small \textsuperscript{2}Baltic Film, Media and Arts School, Tallinn University\\  
\small ‡Corresponding authors: varvarag@ust.hk, mar.canet@tlu.ee \\
\small *equal contribution as first authors  
}

\date{\small June 8, 2023 \vspace{-0.4cm}} %
\begin{document}
\maketitle

\begin{abstract}
Artificial Intelligence is present in the generation and distribution of culture. How do artists exploit neural networks? What impact do these algorithms have on artistic practice? Through a practice-based research methodology, this paper explores the potentials and limits of current AI technology, more precisely deep neural networks, in the context of image, text, form and translation of semiotic spaces. 
In a relatively short time, the generation of high-resolution images and 3D objects has been achieved. There are models, like CLIP and text2mesh, that do not need the same kind of media input as the output; we call them translation models. Such a twist contributes toward creativity arousal, which manifests itself in art practice and feeds back to the developers’ pipeline. Yet again, we see how artworks act as catalysts for technology development.
Those creative scenarios and processes are enabled not solely by AI models, but by the hard work behind implementing these new technologies. AI does not create a ‘push-a-button’ masterpiece but requires a deep understanding of the technology behind it, and a creative and critical mindset. Thus, AI opens new avenues for inspiration and offers novel tool sets, and yet again the question of authorship is asked.
\end{abstract}

\begin{figure}[t]
         \centering
         \includegraphics[width=\linewidth]{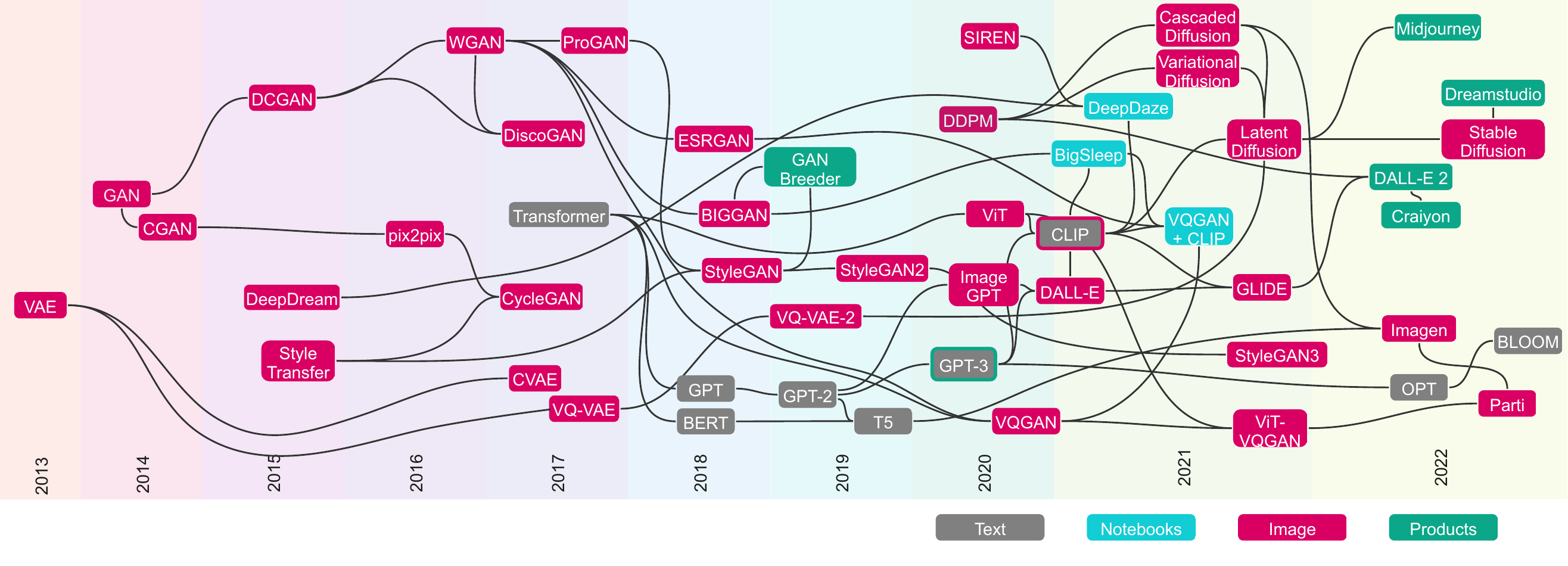}
         \caption{Timeline of creative deep learning development.}
  \label{fig_1}
\end{figure}

\section{Introduction}

It is claimed that recent advancements in AI, such as CLIP-based products Midjourney and DALL-E, are supposed to augment our creativity. For the first time, it does not sound so absurd that artists can find themselves out of jobs \parencite{nicholas2017these}. Not that artists would have ever had a secure and stable job, but deep learning (DL) tools might eventually lead to losing some commercial commissions. Such thinking relies on a modern art approach where skills are in the centre of attention and not the conceptual idea. Quoting Lev Manovich: “Since 1970 the contemporary art world has become conceptual, ie focused on ideas. It is no longer about visual skills but semantic skills.” \parencite{manovich2022ai} As these new tools advance, the interfaces and techniques become more complex and sophisticated as our eyes are becoming more accustomed to not being easily surprised. 

Echoing Aaron Hertzmann, once painters were in a similar situation when photography was invented and took over the niche of portrait-making. Then visual artists had to re-invent themselves and re-think the meaning of painting. Photography had to wait another 40 years until it got recognized as an artistic medium \parencite{hertzmann2018can}. So-called AI artists have faced similar challenges in gaining acceptance within the art world and even inside the digital art niche \parencite{roose2022ai}. 

Computer art emerged with the invention of the computer. Artists, such as Vera Molnar and Manfred Mohr, created their first computer-generated artworks in the 1960s using scientific lab computers at night when they were not used by scientists. Early computer artists were re-purposing a machine for artistic use and writing code to make art on it. Since the creation process was mediated by a computer, it may seem to the general audience that the artists were simply pressing a button and the computer doing art for them. Hence, the question of authorship emerged: is the artist a machine or human?

Today, with the appearance of neural networks (NN) and their creative applications, the same question re-appears. Hertzmann has written several articles arguing that people do art and not computers \parencite{hertzmann2018can,hertzmann2020computers}. Manovich also describes how AI-generated images that imitate realist and modernist paintings are claimed to be art \parencite{manovich2022ai}. At the same time, experimental art forms, like installation, interactive format, performance and sound art, are often overlooked unless they are promoted by a large corporation.
Instead of re-telling a short but very dense history of DL technology development, in the next section, we focus on the appearance of neural network tools that raised interest amongst artists and led to meaningful artwork production.

\section{Historical overview of DL development}

DL is a subset of machine learning (ML) using Deep Neural Networks (DNN) to learn underlying patterns and structures in large datasets. In 2012, a DNN designed by Alex Krizhevsky outperformed other computer vision algorithms to achieve the new state of the art in the ImageNet Large Scale Visual Recognition Challenge \parencite{heravi2016classification}. This model, AlexNet, signalled the start of a new DL era. As AI technology has developed and become more prevalent in real-world systems, artists have been exploring its limits and potentials, adapting these models to their own practices. As the number of scientific publications on AI grows exponentially it is useful to map out the influential papers, and related applications, to help track the evolution of the AI-Art space in relation to the technological advances \parencite{krenn2022predicting}. Figure \ref{fig_1} shows a timeline of the development of generative models for images and text. Using this diagram we can make a few observations on the past ten years: the dominance of GANs for image generation, the influence of the Transformer on 

Large Language Models (LLM), and the growing interest in multi-modal approaches and translation models.
The starting period of image generation using DNNs can be traced back to the creation of the Variational Auto-Encoder (VAE) in 2013, and the Generative Adversarial Network (GAN) in 2014 \parencite{kingma2013auto, goodfellow2020generative}. These models showed different ways in which a NN can be trained on a large dataset, and then used to generate outputs that resemble but do not copy the original dataset. 

For much of the past decade, GAN art has been a dominant and defining element of AI Art. GANs are trained using a competitive lying game, played by two players: the Generator and the Discriminator. The Generator wins by making an image that the Discriminator thinks is from the original dataset. The Discriminator wins by successfully identifying which images the Generator has made. By playing this game repeatedly, both sides slowly learn when they have been fooled and remember information so they don’t fall for the same tricks again. The Generator gets better at making images, and the Discriminator gets better at detecting these fakes. At the end of the game we are left with a Generator that is very good at generating new images, with the qualities and style of our original inputs.
After the original GAN paper, there was a rush of exploration of this new technique for generating images. Alongside general improvements to the models architecture and stability, new ways of guiding the outputs and applying GANs to specific problems were also explored \parencite{radford2015unsupervised,arjovsky2017wasserstein}.

Image-to-Image Translation with Conditional Adversarial Nets (2016), also known as pix2pix, showed a process of converting one type of image into another type \parencite{isola2017image}. Mario Klingemann’s work \textit{Alternative Face}\footnote{\url{https://underdestruction.com/2017/02/04/alternative-face/}} used the pix2pix model with a dataset of biometric face markers and the music videos of the singer François Hardy. This allowed him to control the movement of the face with this form of digital puppetry, which he then demonstrated by transferring the facial expressions of the political consultant Kellyanne Conway onto Hardy’s face as she talks about “alternative facts”. 

In 2015, on the Google research blog, the post Inceptionism: Going Deeper into NNs described a tool that attempted to understand how image features are understood in the hidden layers of the NN \parencite{mordvintsev2015inceptionism}. Alongside this post they released a tool called DeepDream. This model enhances an image with the NN's attempts to find the features of the dataset it was trained on. The creative use of DeepDream was proposed by the authors in the original article “It also makes us wonder whether neural networks could become a tool for artists—a new way to remix visual the creative process in general”.[13]

DeepDream’s psychedelic imagery quickly caught the attention of the internet and of artists around the world, resonating with those interested in understanding the cross-over between biological and neurological construction of images. Memo Atken’s work \textit{All Watched Over By Machines Of Loving Grace}\footnote{\url{https://www.memo.tv/works/all-watched-over-by-machines-of-loving-grace-deepdream-edition/}}: Deepdream edition, hallucinated over an aerial photograph of the GCHQ headquarters. This work raises questions around the motivations of the organisations funding the development of AI, and in doing so make the dreamlike qualities a little more nightmarish. 

In the same year, the paper A Neural Algorithm of Artistic Style introduced a DNN “to separate and recombine content and style of arbitrary images, providing a neural algorithm for the creation of artistic image” \parencite{gatys2015neural}. Neural Style Transfer (later known simply as StyleTransfer) takes two inputs, a style image and a content image, it extracts textural information from the style image and compositional information from the content image, then generates an image with minimal distance between the two. The paper demonstrates this with images of a DeepDream’s psychedelic imagery quickly caught the attention of the internet and of artists around the world, resonating with those interested in understanding the cross-over between biological and neurological construction of images. Memo Atken’s work All Watched Over By Machines Of Loving Grace: Deepdream edition, hallucinated over an aerial photograph of the GCHQ headquarters. This work raises questions around the motivations of the organisations funding the development of AI, and in doing so make the dreamlike qualities a little more nightmarish. 

In the same year, the paper A Neural Algorithm of Artistic Style introduced a DNN “to separate and recombine content and style of arbitrary images, providing a neural algorithm for the creation of artistic image” \parencite{gatys2015neural}. Neural Style Transfer (later known simply as StyleTransfer) takes two inputs, a style image and a content image, it extracts textural information from the style image and compositional information from the content image, then generates an image with minimal distance between the two. The paper demonstrates this with images of a photograph represented in various styles of famous paintings, such as Van Gogh’s \textit{The Starry Night}.

In 2017, CycleGAN continued with the problem of image-to-image generation shown in pix2pix, but removed the requirement of aligned image pairs being needed for training \parencite{zhu2017unpaired}. Instead a set of source images and a set of target images that are not directly related can be used. The advantage of this is it is simpler to scale to larger datasets, making the process more accessible for artists. Helena Sarin has been using CycleGAN for a number of years, and recently in \textit{Leaves of Manifold}\footnote{\url{https://www.nvidia.com/en-us/research/ai-art-gallery/artists/helena-sarin/}}\footnote{\url{https://twitter.com/NeuralBricolage/status/954027624728354821}} she collected and photographed thousands of leaves to build her own training dataset, and then implemented a custom pipeline with changes that improve results when working with smaller datasets. This personalised approach in crafting the models resonates with the hand-made, collaged aesthetic of the images generated.

Other notable developments to GANs brought improvements to image quality and resolution \parencite{karras2017progressive,wang2018esrgan}. In late 2018, the release of StyleGAN, a model built on a combination of ideas from Style Transfer and PGGAN, demonstrated very convincing images of human faces \parencite{karras2019style}. In his article “How to recognize fake AI-generated Images”, the artist Kyle McDonald investigated the images generated by StyleGAN, and highlighted the visual artefacts he found \parencite{mcdonald2018recognize}. At a glance these images look like photographs, but on closer inspection irregularities such as patches of straight hair, misaligned eyelines, or mismatched earrings reveal the difficulties GANs have in managing “long-distance dependencies” in images.

In 2017 the paper Attention Is All You Need proposed a new network architecture called the Transformer \parencite{vaswani2017attention}. This model addressed the long-distance dependency issue in RNNs and CNNs by rethinking how we could handle sequences. Rather than looking at a sentence word by word, the Transformer observes the relationship between all elements of the sequence simultaneously. Being able to better handle long distance dependencies meant the Transformer was appropriate for natural language generation. Artists have explored the use of VAEs for short text generation, but with the emergence of LLM passages of long, coherent, texts could be generated \parencite{brown2020language}. As dataset sizes increased, along with hardware costs for training these large models, they have become harder for individuals to train themselves, and the mode of interaction has shifted from curated datasets and homemade scripts, to web APIs and third party services. While it is more difficult to participate in the training process, the availability of services and interfaces provides new ways of working with these models that can produce less technical and more playful approaches. For example, Hito Steyerl used GPT-3 to create Twenty-One Art Worlds: A Game Map and described the process as “fooling around” with GPT-3 to write descriptions of different Art Worlds \parencite{steyerl2022twentyone}. In the resulting text it is difficult to distinguish which words may have been written by Steyerl and which were written by GPT-3.
    
The learnings from LLM for text generation were soon applied to image generation (Image GPT, Vision Transformer), and the simultaneous release of CLIP and DALL-E in January 2021 signalled the start of a new era of image generation \parencite{chen2020generative,dosovitskiy2020image}. Although the DALL-E model was not released, CLIP was made available to the public, and the model was quickly adopted by AI artists who applied the idea of CLIP guidance to various image generation techniques. Ryan Murdock produced the colab notebooks DeepDaze\footnote{\url{https://github.com/lucidrains/deep-daze}} (combining CLIP and SIREN) and BigSleep\footnote{\url{https://github.com/lucidrains/big-sleep}} (CLIP and BIGGAN), which were subsequently adapted by Katherine Crowson in the widely distributed VQGAN+CLIP\footnote{\url{https://github.com/EleutherAI/vqgan-clip}} notebook.

The paper Denoising Diffusion Probabilistic Models  introduced a different method for creating generative models \parencite{ho2020denoising}. This technique trains a model by adding increasing amounts of noise to an image and then having the model remove the noise, resulting in a model that can generate images from only noise. Diffusion models, when combined with CLIP or other conditioning processes, enable much faster text-to-image processing. The popularity and accessibility of these techniques was further raised by the release of DALL-E 2 and Midjourney in 2022. Midjourney became so popular it is now the largest Discord server with over 5 million members. Following the releases of these products, open source models such as Stable Diffusion have also been developed. There are many benefits of using free and open source models for artists. Being able to modify code and develop on your own software allows the artist to pursue their own experimental approaches, not restricted to the interface designed by a service provider. 

The artist's involvement in generating new images with these models is vastly different to working with GANs. Rather than building custom datasets and training models, instead the focus has shifted to writing prompts that can generate the images the artist wants to find, and designing interfaces for exploring these prompts and their translations. The artist Johannez coined the term Promptism for describing his art practice, and wrote a humorous Prompist manifesto using GPT-3. Against a backdrop of models trained on hundreds of millions of images scraped from the internet, including many artists’ portfolios, the manifesto asserts “The prompt must always be yours” \parencite{johannezz2022promptist}.

\section{Artist-Guided Neural Networks}

Many papers discuss AI from the point of view of creativity taking mostly one position of two: either AI as an amazing tool for artists and creativity, or AI is seen as something negative in art. It is easy to see that the people from industry advocate for the first position, and theory scholars for the second one. But, how do practitioners see contemporary AI technology themselves? And in which ways AI is deployed in art practice? Hence, it is not the focus of this paper to discuss whether AI can make art, but rather how AI can be useful for artists and what new ideas it can offer. By using practice-based research methodology, we decode the role of AI tools in artistic practice and trace the evolution of such artistic work. In this paper, the practice of artist duo Varvara \& Mar was used as a case study, which provided us with the insides in this research. We divide the case studies into four categories based on medium: synthetic image, synthetic text, synthetic form, and translation models. From the view of the practitioner, the limitations, new possibilities, and change in production processes are discussed.

\begin{figure*}[t]
 	\noindent
 	\includegraphics[width=\textwidth]{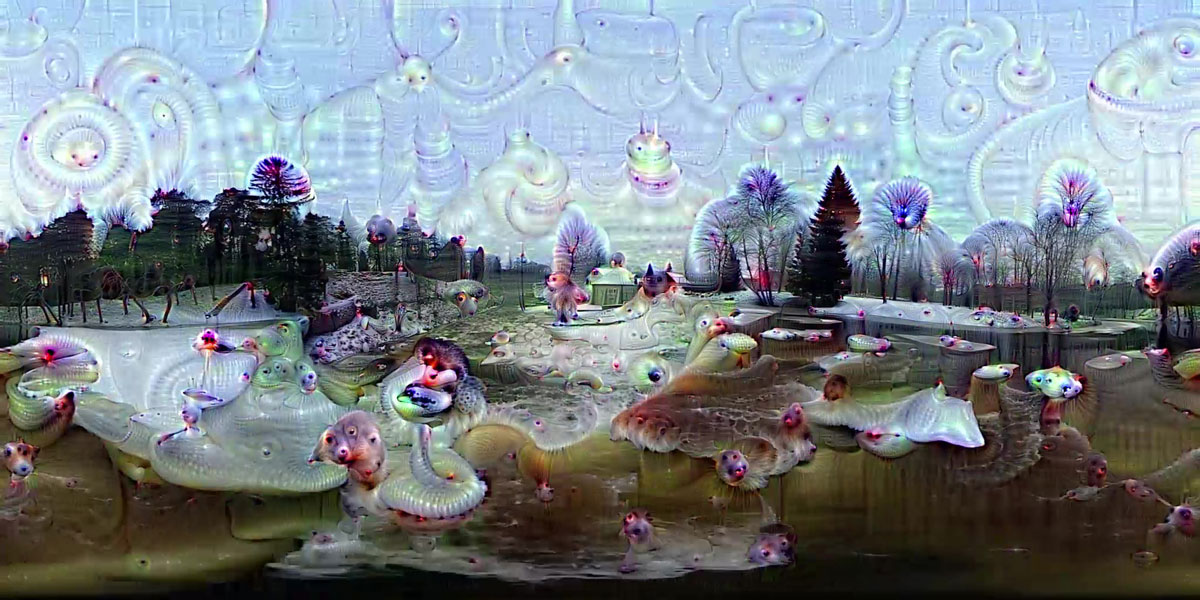}
 	\caption{ A single still image from the VR 360º video \textit{Neural Landscape} (2017). ©Varvara \& Mar.}
  \label{fig_4}
\end{figure*}

\begin{figure*}[t]
 	\noindent
 	\includegraphics[width=\textwidth]{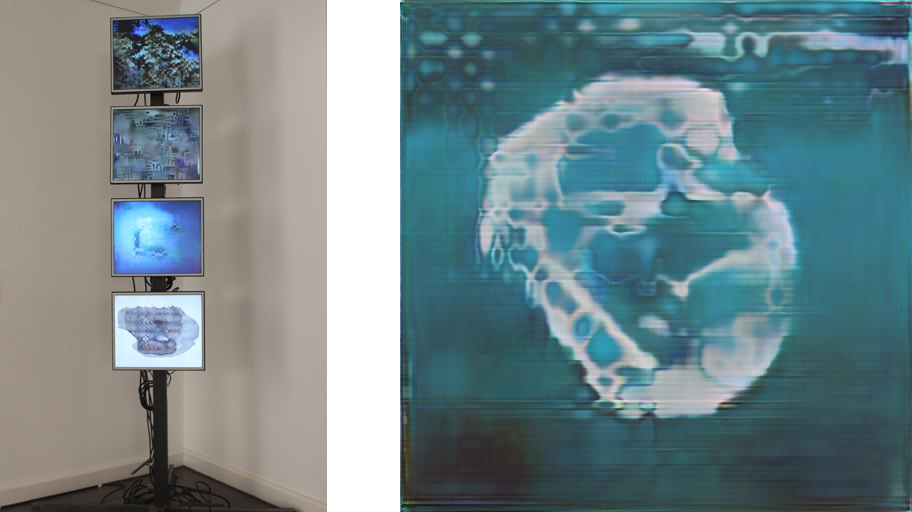}
 	\caption{ Left: installation view of \textit{Plasticland} (2019). Right: an AI-generated image from the dataset of platic under the water. \textit{Plasticland} (2019). ©Varvara \& Mar.}
  \label{fig_Plasticland}
\end{figure*}

\begin{figure*}[t]
 	\noindent
 	\includegraphics[width=\textwidth]{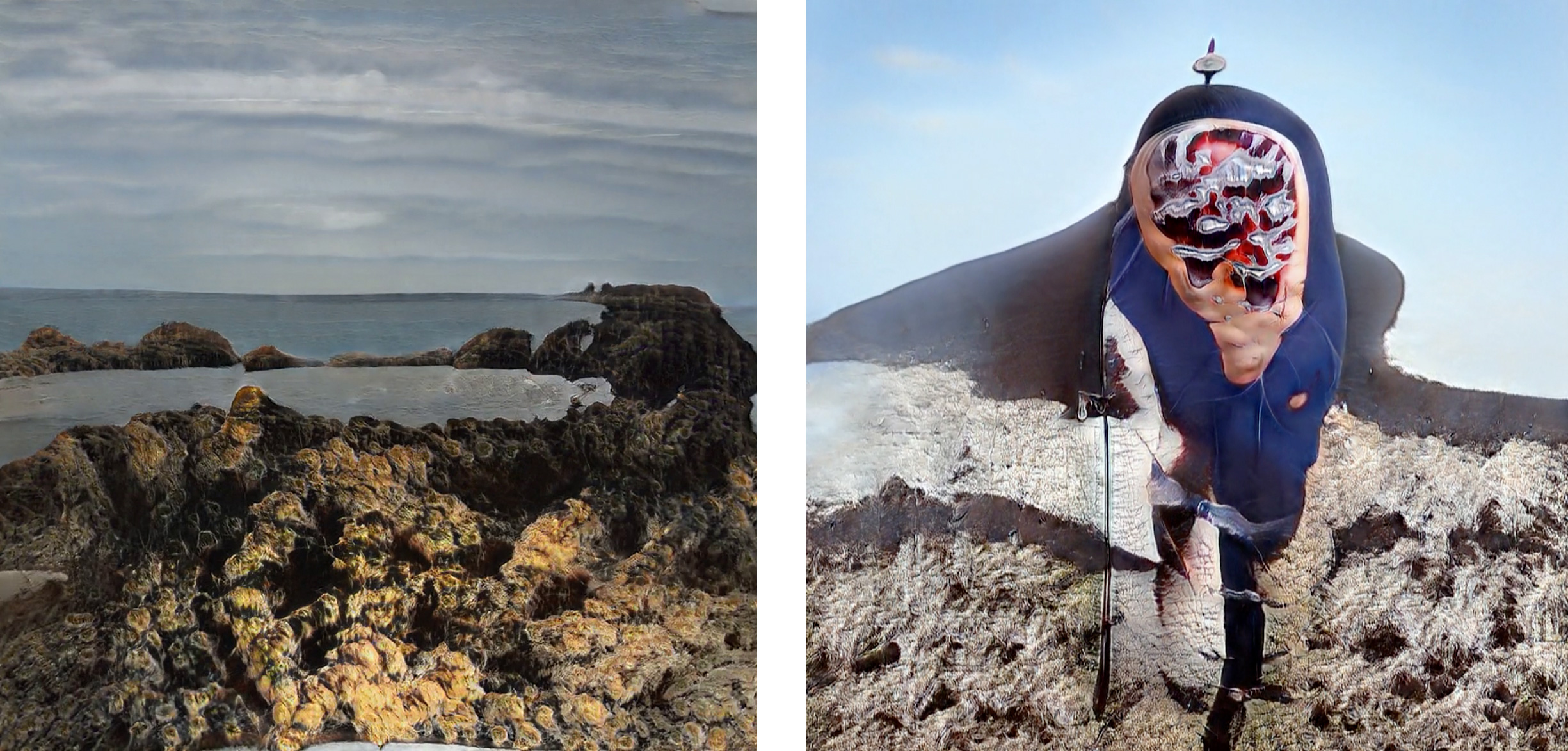}
 	\caption{ Single still images from the two AI-generated videos \textit{POSTcard Landscapes from Lanzarote I,II} (2020). ©Varvara \& Mar.}
  \label{fig_POSTcard}
\end{figure*}

\begin{figure*}[t]
 	\noindent
 	\includegraphics[width=\textwidth]{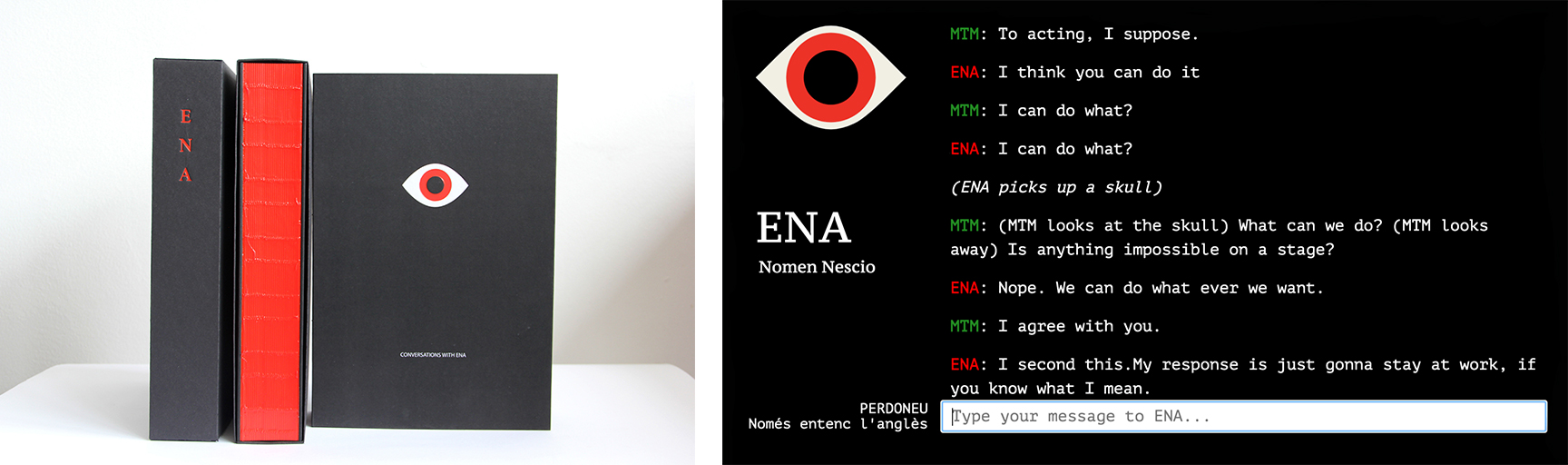}
 	\caption{ \textit{ENA} (2020). Left: ENA Book with all conversations. Right: Screenshot of the website app in the Theatre Lliure installed during May 2020 during Covid lock down times.  ©Varvara \& Mar.}
  \label{fig_ENA}
\end{figure*}

\subsection{Synthetic Image}

Our DL exploration began in 2017 with Google DeepDream, focusing on image generation (Fig.\ref{fig_4}). The concept behind \textit{Neuronal Landscapes}\footnote{\url{https://var-mar.info/neuronal-landscapes/}} project was to imagine how Estonian landscape will look like in 100 years time (commission work for the Estonian History Museum). Through synthetic vistas created by machines, the artwork offers a glimpse into the environment from a machine's perspective, immersing viewers in a hallucinated neural net simulacrum. To depict the evolution of Estonian society over time, from forests and farmlands to urbanization and digitalization, a 360º VR video was created. Filmed with drone-mounted two 360º cameras, the footage was edited and processed using DeepDream. The rendering process spanned 30 days on powerful machines with Nvidia TitanX GPUs. While some customization was possible, the algorithm's aesthetic footprint remained prominent.

In the next art project, ProGAN was deployed. For the first time we worked with datasets and training GAN models. \textit{Plasticland}\footnote{\url{https://var-mar.info/plasticland/}} (2019) talks about plastic waste and ecological problems this material causes (Fig.\ref{fig_Plasticland}). We composed four different datasets of images of layered plastics in our planet: landfills, plastic on top of water, plastic underwater, and plastiglomerates. The ProGAN model was trained on a local machine using pyTorch and took a week to train, and the artist used a selection of generated images to create a video composition. A metal totem displaying those synthetic, as plastic is, layers, we draw attention not only to the problem of waste but also question whether AI has some similarity with this material. Since the invention of plastic, this material was applied almost everywhere because of its perfect qualities, until we realised that it is not sustainable and ecology-friendly. Will a similar story happen with AI? From the practice-based research perspective, this work shows artists’ desire to move from a still to moving image and towards sculptural form that is held back by the early stage of machine learning technology: low resolution images jumping from one frame to another.

The next artworks \textit{POSTcard Landscapes from Lanzarote I} (00:18:37) and II (00:18:40)\footnote{\url{https://var-mar.info/postcard-landscapes-from-lanzarote/}} in 2021 demonstrate the artist's ability to create video works with StyleGAN2 (Fig.\ref{fig_POSTcard}). The hypnotic appearance of these works, where one frame morphs naturally into another, shows the artists' ability in guiding the outputs of the neural network. Vector curation and composition of a journey through the latent space, created by training the model on specific datasets of 2000+ images, were crucial and integral parts of the artistic process. The artwork talks about critical tourism and how circulation of images representing touristic gaze overpower the nature of seeing. In the words of Jonas Larsen “‘reality’ becomes touristic, and item for visual consumption” \parencite{larsen2006geographies}. Hence, we scraped, where licence allowed, the location-tagged images from Flickr and composed two datasets of photos categorised  as tourism or landscape. As we have written earlier: “The two videos are random walks in the latent space of the Stylegan2 trained models, creating a cinematic synthetic space. The audiovisual piece shows an animated image through the melted liquid trip of learning acquired from the dataset composed of static images. The video flows from point to point, generating new views and meaning spaces through the latent space’s movement. The audio was created after the video was generated in response to the visual material to complete the art piece.” \parencite{guljajeva2022postcard}. The sound for local or landscape view was created by a sound artist from Lanzarote, Adrian Rodd, who aimed to give a socio-political voice to the piece. In contrast, the sound design created by Taavi Varm is a soundscape replying to touristic gaze. The artists aimed to initiate collaborations with others but also to experiment with human-AI co-creation. In a similar vein is the artwork Phantom Landscapes of Buenos Aires (00:20:00, 2021), with sound work by Cecilia Castro.

Our last experiment with GAN models \textit{Synthetic-scapes of Tartu} (00:10:00, 2022), demonstrates a different approach. Taking a dataset composed from our own video footage (flaneur walks), we first produced the sound (a composition by Taavi Varm, Ville MJ Hyvönen with piano by J. Kujanpää) and used this to inform the direction of the video . The result was a sound-guided AI-generated visual output. 

\subsection{Synthetic Text}

In this section, we focus on artwork incorporating AI text generation as part of the artistic concept. Our journey to text generation started with the online participative theatre project \textit{ENA}\footnote{\url{https://var-mar.info/ena/}} and ended with a hand-bound publication (Fig.\ref{fig_ENA}).
    
During the first lockdown in May 2020, together with theatre maker Roger Bernat, we created an online participative theatre piece \textit{ENA} on the website of Theater Lliure in Barcelona. \textit{ENA} is a generative chatbot that talks to its audience, and together (AI and audience), they make theatre. As we have described before: “Although in the description of the project it was stated explicitly that people were talking to a machine, multiple participants were convinced that on the other side of the screen another human was replying to them—more precisely the theatre director himself, or at least an actor.” \parencite{guljajeva2021ena}.

Analysing synthetic books, Varvara Guljajeva has stressed the importance of human input in the AI text-generation systems \parencite{guljajeva2021synthetic}. In addition, one also needs to guide the audience participation and interaction with the chatbot. For this purpose, we have adopted the traditional theatre method for guiding actors, as a way to guide the audience, and thus, the bot, too. Stage directions were used as a guiding method, which triggered thematic conversation and offered meaningful dialogue between humans and the AI system. We found the conversations so meaningful that we decided to publish a book that contains all the conversations with \textit{ENA}.

With this project, we learned that it is essential to guide neural networks via audience interaction. In order to do this, it is also necessary to guide the audience. Without audience interaction guidance, it is nearly impossible to achieve meaningful navigation of neural networks.

\begin{figure*}[t]
 	\noindent
 	\includegraphics[width=\textwidth]{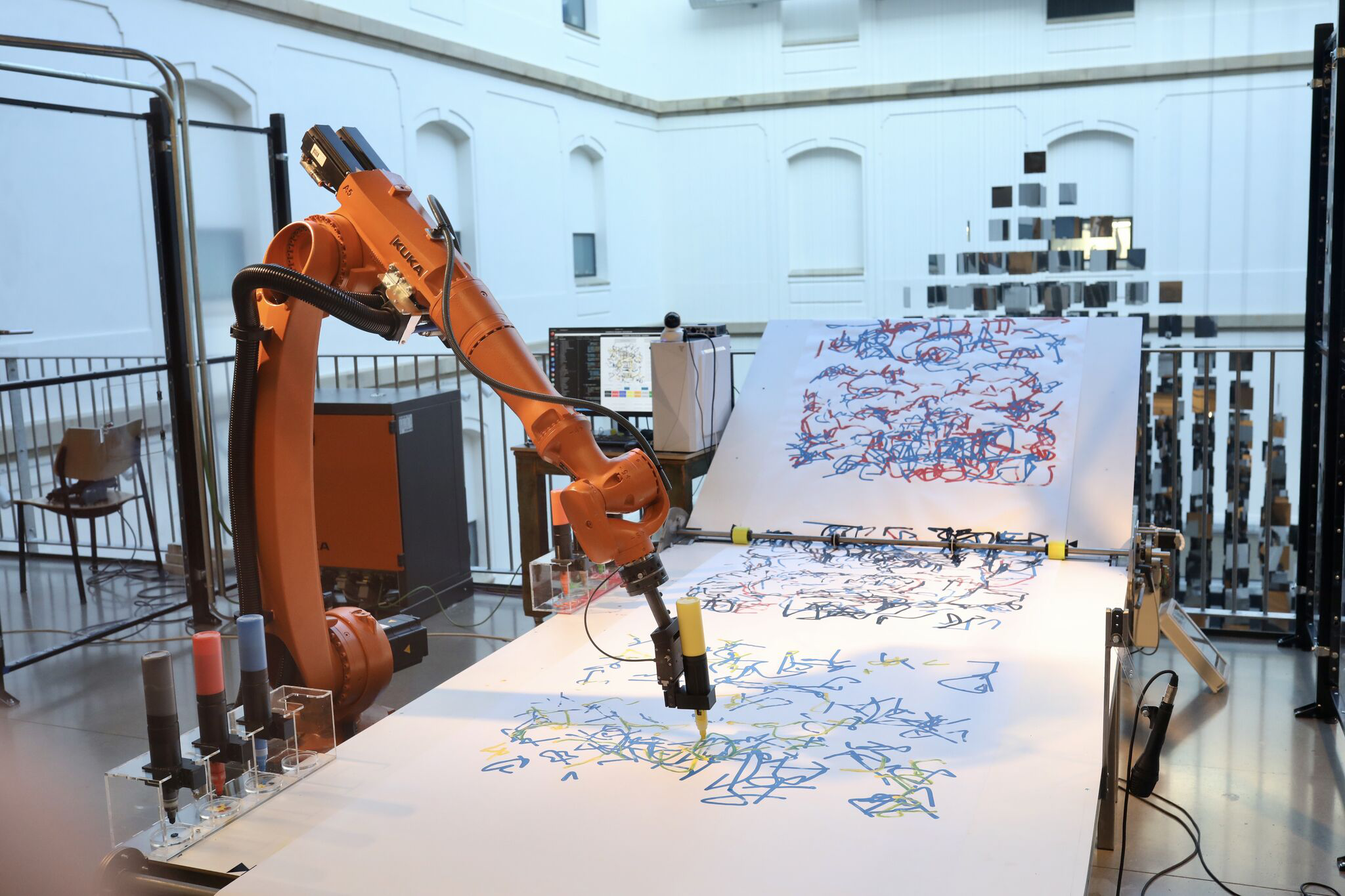}
 	\caption{ Kuka industrial robot painting audience’s dreams. Installation view of Dream Painter (2021). ©Varvara \& Mar.}
  \label{fig_2}
\end{figure*}

\subsection{Translation models}

This category focuses on translation models that enable interactive and installation-based formats. Translation refers to the conversion of mediums, or as we put it, translation of semiotic spaces. To illustrate this, we introduce \textit{Dream Painter}\footnote{\url{https://var-mar.info/dream-painter/}} an art installation that translates audience’s spoken dreams to a line-drawing produced by a robot (Fig.\ref{fig_2}). As described earlier: “\textit{Dream Painter} is an interactive robotic art installation that explores the creative potential of speech-to-AI-drawing transformation, which is a translation of different semiotic spaces performed by a robot. We extended the AI model CLIPdraw which use CLIP encoder and the differential rasterizer diffvg for transforming the spoken dreams into a robot-drawn image.” \parencite{canet2022dream}. “Design- and technology-wise, the installation is composed of four larger parts: audience interaction via spoken word, AI-driven multi-colored drawing software, control of an industrial robot arm, and kinetic mechanism, which makes paper progression after each painting has been completed. All these interconnected parts are orchestrated into an interactive and autonomous system in a form of an art installation […].” \parencite{guljajeva2022dream}. Out of all the projects discussed, this was the most difficult to realise. This is because of the large scale of the artwork, and multiple parts of software and hardware that need to run automatically and synchronously.

In this project we investigated how guidance of neural networks could be interactive and real-time instead of non-interactive and pre-determined, as shown in previous examples of our work. It is important to notice that methods, such as dataset composition and output curation were not used in this case. In fact, visual output curation is totally missing. The artists created an interactive system to be experienced and discovered by the audience. This means the audience determines the output. Instead of curating a dataset, a CLIP model is used that can produce nearly real-time output guided by a text prompt. As we have written earlier: “Translation of semiotic spaces, such as spoken dreams to AI-generated robot-drawn painting, allowed us to deviate from image-to-image or text-to-text creation, and thus, imagine different scenarios for interaction and participation.” \parencite{guljajeva2022dream}.

This project indicates our search for transformative outputs of AI technology, and thus, shows the evolution in practice. By extending available DL tools and combining with other technology, for example, text-to-speech models, real-time industrial robot control, and physical computing, it offered an interactive robotic and kinetic experience of neural network latent space navigation. This contributes towards the explainability of AI because the audience could experience how the words affected the drawing, and which concept triggered which outcome.

Being inspired by Sigmund Freud's work on the interpretation of the human mind while unconscious, we speculatively ask if AI is powerful enough to understand our dreamworld. Through practice we question the capacities of neural networks and investigate how far we can push this technology in the art context. This artwork allows the audience to experience the limits of concept-based navigation with AI. The system is unable to interpret and can only illustrate our dreams. It cannot understand the prompt semantically and only gets the concepts.

\begin{figure*}[t]
 	\noindent
 	\includegraphics[width=\textwidth]{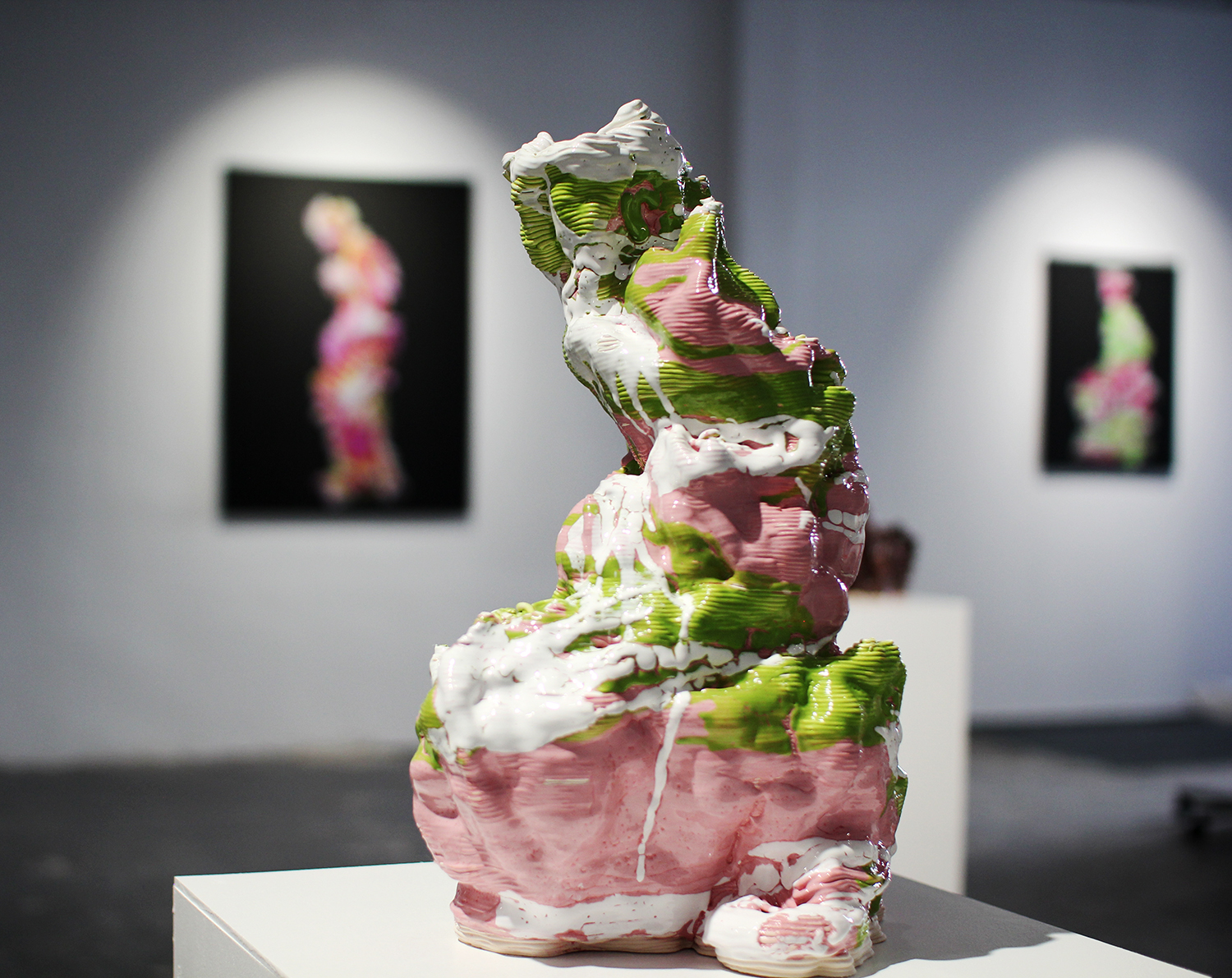}
 	\caption{Ceramic sculpture guided by 3D object and text prompt, 3D printed in clay, and glazed manually. This piece belongs to the series Psychedelic Forms (2022). ©Varvara \& Mar.}
  \label{fig_3}
\end{figure*}
 
\subsection{Synthetic Form}

In this section, we ask how artists can guide neural networks when creating volumetric forms, and what happens when AI meets materiality. After working for a while with DL tools that produce 2D outputs, it is an obvious step to explore possibilities to produce 3D results. To our surprise, it was not an easy task to find the solution (Oct 2021). \textit{Psychedelic Forms} is a series of sculptures produced in ceramics and recycled plastic through which we investigated the possibilities of AI in producing physical sculptures. The project re-interprets antique culture in the contemporary language and tools \parencite{guljajeva2023ai}.
Following the same paradigm shift as in the previous section, text2mesh is a CLIP-based model that does not require a dataset, but a 3D object and text prompt as input \parencite{michel2022text2mesh}. Hence, the model actually does not create a 3D model but stylises the inserted one, guided by inputted text.

We decided to go back to the origins, in terms of ancient sculptures and material selection. Although it was said that there was no dataset, we still had a collection of 3D models of ancient sculptures because, by far, not all produced a desirable output. In this sense, there was definitely an output curation present in the process.
The criteria for selection were the following: first, the form had to be intriguing, and second, it should be possible to produce it in material afterwards. It was clear that we had to modify each model because the physical world has gravity, and the DL model does not take this into account. Some generated models were discarded because they were seen as not-fixable, although interesting in their shape.

The process demonstrated here is quite an unusual way to create an object. After extensive experimentation with the tool, we learned how certain words triggered certain shapes and colours. This knowledge gave us a chance to treat text prompts as poetic input. Thus, we created short poems to guide NN. The best ones survived as titles and are reflected in the forms.
The artists did not strictly follow the original model but took the creative liberty to modify the shape and determine the colour by manually glazing the sculptures. The dripping technique was used for colouring the sculptures.  This served as a metaphor for liquid latent space and the psychedelic production process (this was the artists’ inner feeling about the creative process because they did not know what results would be achieved in the end). Sometimes, AI-generated vertex colouring was taken as inspiration, sometimes totally ignored. Nevertheless, digital sculptures were exhibited alongside the physical ones to underline the transformation and human role in the creative process. Although ceramic sculptures were 3D printed in clay, the fabrication process had to follow the traditional way of producing pottery (Fig.\ref{fig_3}). Since the artists had never engaged in ceramics before, the whole production process felt psychedelic: unexpected neural network processes led to transformation by numerical, physical, and chemical processes, all guided by both the artists and chance. Hence, the art project highlights the relationship between different agencies.

In the end, we can say that AI is not prepared for the physical world. It created nice images, but when one wants to materialise the output, it requires considerable additional work. However, those extra processes were very rewarding and creative in our case. In this project, AI served as an inspiration or a departing point more than anything else. In 
other words, the experimental phase of technology is necessary for experimental practices, and this can lead to the creation of a new production pipeline. The fine line between control and chance when guiding the neural networks and related processes is likely the main creative drive for the artists.

\section{Discussion}

According to the media hype around AI, this technology is intelligent enough to create art autonomously \parencite{perez2018microsoft,vallance2022art}. However, the reality is different. A computer scientist and a co-inventor of Siri Luc Julia, AI does not exist. He advocates for machines’ multiple intelligences that often outperform humans. However, machine intelligence is limited and discontinuous compared to human intelligence \parencite{julia2020there}. Therefore, it is vital to have artistic practices around this technology, as a counterbalance to the AI fantasies served by the industry and mass media.

We see AI as a creative tool with its own possibilities and limitations, which can stimulate artists’ creativity through unexpected outputs. Research has shown that tool-making expands human cognitive level and constitutes evolution in culture \parencite{stout2011stone,stout2016tales}. Similarly, as a new tool, generative AI could potentially enrich creativity by allowing new production pipelines that can create unique results.

Coming back to the synthetic images, we can say that all machine-created synthetic image-based works discussed here have particular aesthetics: both with DeepDream and GAN. Unlike the output of GANs, DeepDream has a more recognizable style and can be seen more as a filter that transforms every inputted image instead of learning from the given dataset. Regarding GAN aesthetics, such visual appearance is inherited from two entities to a large extent: the dataset and the model itself. GANs have a particular footprint, as seen in all works produced with this model. The visual palette comes from the used datasets. For example, if a dataset is homogeneous (only landscape images), then we will easily recognize landscapes in the generated output. However, if images in the dataset have a lot of visual variation, the output is rather abstract. \textit{POSTcard Landscapes from Lanzarote II} illustrates this well. Also, when photos in the dataset look similar, the output will also be similar, as was the case with the 
\textit{Synthetic-scapes of Tartu} video work where frames from recorded flaneur walks in a city were extracted. When we talk about video works generated with the neural net, then manual guidance of latent space offered more variations than an audio-led approach.

Synthetic image works have encouraged us to work with formats like images and videos that we did not engage in before in our art practice, but we found it exciting working with AI and video. For example, AI video generation has some affordances, like starting and ending can be done in a perfect loop since images are synthetically generated. However, creating real-time AI work is much more complex because some models are too slow. It might take a few minutes to render a single image. The limitations inspire us to devise new solutions and work in new mediums. Moreover, the limitations of the medium has always been a good challenge for our creativity. 

Working with GANs or other image-generation tools has become much easier in recent years, although it used to be quite difficult. We must note that for practitioners, easy-to-use tools, such as DALL-E and Midjourney, offer little creative freedom, and thus, are less attractive to the artists. Those products tend to instrumentalize the user rather than the other way around. At the same time, open source models offer more creative freedom and enable broader use of artistic ideas. 

The work with generated text demonstrates that AI is not context-aware but maps concepts automatically without understanding semantics. More importantly, as shown in the \textit{ENA} project the audience must also be guided alongside the AI. In the case of \textit{ENA}, stage directions were used, and in the Dream Painter project, the concept of dream telling was applied to guide the participants who in turn guided the neural net through their interaction, creating a chain reaction. Navigating concepts in latent space is artistically interesting and inspiring, this was especially evident when working with form. The artists went beyond semantics and learned how to guide neural networks with a text prompt and 3D object.

The presented practice represents a paradigm shift in machine learning, moving away from composing datasets for GANs and toward translating semiotic spaces enabled by diffusion models. The evolution in practice shows how artists discover and learn to work with the DL toolset, embracing its possibilities and limitations. In the case of practice-based research, practice can be seen as a lab for testing artistic ideas with technology through chance until control is encountered. 

\section{Conclusion}

In this article, we have summarised DL development from the perspective of artists’ interests concentrating on the image, video, text, 3D object generation, and translation models. We applied practice-based research methodology to investigate the role and possibilities of recent co-creative AI tools in artistic practice.

It is difficult to keep pace with AI development. In less than a decade, we have gone from blurry black-and-white faces to impressive high-resolution images guided by text prompts. The user level has gone from difficult to easy, which on one side, broadens possibilities for creation, but on another, it diminishes experimentation and creativity, since AI outputs seem ready-made. This is also demonstrated by the explorative nature of the body of work presented here.
    
Furthermore, it was noticed that creative AI, especially GAN models, have recognizable aesthetics, which in the long run, become repetitive. This led to the change of tools by the artists. The curation of datasets, models, and outputs, along with neural network guidance, have become the toolset of an artist working with AI. Finally, these models can generate multitudes of outputs, but the art is giving the right input to guide the desired output and selecting the results that best serve the concept 
    
As Andy Warhol had envisioned in 1963, eventually, art production will become mechanised and automated \parencite{10.1093/oxartj/kcy001}. In his own words: “I want to be a machine” \parencite{bergin1967andy}, which was also a reflection on that time's vast industrialization process. Resonating with today’s deep learning age: I want my machine to do art.

\section*{Author contributions, acknowledgments and funding} 

 MSC is supported as a CUDAN research fellow and ERA Chair for Cultural Data Analytics, funded through the European Union’s Horizon 2020 research and innovation program (Grant No.810961). 
 
\begingroup
\setlength{\emergencystretch}{8em}
\printbibliography
\FloatBarrier
\endgroup

\end{document}